\documentclass[11pt, letterpaper]{article}

\usepackage{ifthen}

\newcommand{\MyMacro}[4]{
\newboolean{#1}
\setboolean{#1}{#2}
\newcommand{#3}{\ifthenelse{\boolean{#1}}}
\newcommand{#4}{\ifthenelse{\not \boolean{#1}}}
}

\usepackage{ifthen}

\ifthenelse{\isundefined{\MyMacro}}{

\newcommand{\MyMacro}[4]{
\newboolean{#1}
\setboolean{#1}{#2}
\newcommand{#3}{\ifthenelse{\boolean{#1}}}
\newcommand{#4}{\ifthenelse{\not \boolean{#1}}}
}

}

\newcommand{\ifndef}[2]{\ifthenelse{\isundefined{#1}}{#2}{}}

\newcommand{\mydef}[2]{\def#1{#2}}

\newcommand{\nospell}[1]{#1}   %

\usepackage{amssymb}
\usepackage{amsmath}   %

\usepackage{amsthm}    %
\usepackage{latexsym}
\usepackage{amsfonts}
\usepackage{units}     %
\usepackage{psfrag}    %
\usepackage{url}    %

\usepackage[dvips]{graphicx}
\usepackage[usenames,dvipsnames]{color}

\newcommand{\MyComment}[1]{\ClassWarning{My Macros}{#1}}

\ifndef{\theorem}{\newtheorem{theorem}{Theorem}[section]}
\ifndef{\lemma}{\newtheorem{lemma}[theorem]{Lemma}}
\ifndef{\corollary}{\newtheorem{corollary}[theorem]{Corollary}}
\ifndef{\remark}{\theoremstyle{remark} \newtheorem{remark}{Remark}}
\ifndef{\proposition}{\newtheorem{proposition}{Proposition}}
\newtheoremstyle{mydefinition}   %
{\topsep}{\topsep}   %
{\slshape}   %
{}   %
{\bfseries}   %
{.}   %
{ }   %
{}   %
{\theoremstyle{mydefinition}}
\newtheoremstyle{myexample}   %
{\topsep}{\topsep}   %
{\itshape}   %
{}   %
{\slshape}   %
{:}   %
{ }   %
{\ul{\thmname{#1}}}   %
\ifndef{\example}{\theoremstyle{myexample} \newtheorem{example}{Example}}
\ifndef{\claim}{\newtheorem{claim}[theorem]{Claim}}
\ifndef{\result}{\newtheorem{result}[theorem]{Result}}
\ifndef{\problem}{\newtheorem{problem}{Problem}}
\ifndef{\protocol}{\newtheorem{protocol}{Protocol}}
\newtheoremstyle{myclaims}   %
{\topsep}{\topsep}   %
{\slshape}   %
{}   %
{\bfseries\itshape}   %
{}   %
{ }   %
{\thmname{#1}\thmnumber{ \!#2}.}   %
{\theoremstyle{myclaims}

\ifndef{\fact}{\newtheorem{fact}[theorem]{Fact}}
\ifndef{\observation}{\newtheorem{observation}[theorem]{Observation}}
}

\newtheoremstyle{anystatement}{\topsep}{\topsep}{\itshape}{}{\bfseries}{.}{ }{\anystatementname}
{\theoremstyle{anystatement}}

\newcommand{\anystatementname}{}

\newcounter{tmp_id_cnt}
\newcommand{\AuxNew}[4][]{#2{#3}[1][*]%
{\ifthenelse{\equal{*}{##1}} %
{\Ensuremath{#1{#4}}}%
{\ifthenelse{\equal{b}{##1}} %
{\Ensuremath{\mathbf{#4}}}%
{\ifthenelse{\equal{}{##1}} %
{\IfMathMode{#1{#4}}{#4}}{}}}}}

\newcommand{\newident}[3][*]{\ifthenelse{\equal{*}{#1}}%
{\AuxNew[\mathit]{\newcommand}{#2}{#3}} %
{\mydef{#2}{\Ensuremath{\mathit{#3}}}}} %

\newcommand{\newidentarg}[2]{%
\newcommand{#1}[1][]%
{\Ensuremath{\mathit{#2}}}}              

\newcommand{\newmat}[3][*]{\ifthenelse{\equal{*}{#1}}%
{\AuxNew{\newcommand}{#2}{#3}} %
{\mydef{#2}{\Ensuremath{#3}}}} %

\newcommand{\providemat}[3][*]{\ifthenelse{\equal{*}{#1}} %
{\AuxNew{\providecommand}{#2}{#3}} %
{\mydef{#2}{\Ensuremath{#3}}}} %

\newcommand{\providematarg}[2]{ %
\providecommand{#1}[1][]{\Ensuremath{#2}}}      

\newcommand{\newmatop}[2]{\mydef{#1}{\operatorname{#2}}}

\newcommand{\MyMakeTheoMacros}[3]{
\newcommand{#2}[2][]{\ifthenelse{\equal{}{##1}}
{\begin{#1} ##2 \end{#1}}
{\begin{#1}\label{##1} ##2\end{#1}}}
\newcommand{#3}[3][]{\ifthenelse{\equal{}{##1}}
{\begin{#1}{\e{##2}} ##3 \end{#1}}
{\begin{#1}{\e{##2}}\label{##1} ##3\end{#1}}}
}

\newcommand{\MyMakeDupTheoMacros}[8]{
\MyMakeTheoMacros{#1}{#2}{#3}
\newcommand{#4}[3]{
\newcommand{##2}{##3}
\begin{#1}\label{##1} ##2\end{#1}}
\newcommand{#5}[4]{
\newcommand{##2}{##4}
\begin{#1}{\e{##3}}\label{##1} ##2\end{#1}}
\newcommand{#8}[2]{\def\my_tmp_id{my_tmp_id_\arabic{tmp_id_cnt}}
\newtheorem*{\my_tmp_id}{#7~\ref{##1}}
\begin{\my_tmp_id} ##2 \end{\my_tmp_id}\stepcounter{tmp_id_cnt}}
\newcommand{#6}[6]{
#2[##1]{##2}

##3
\prf[#7~\ref{##1}]{##6} \newcommand{##5}{}

}
}

\newcommand{\MyMakeRefMacros}[3]{\newcommand{#1}[2][]
{\ifthenelse{\equal{}{##1}}{#2~\ref{##2}}{#3~\ref{##1} and~\ref{##2}}}}

\newcommand{\MyMakeEqRefMacros}[3]{\newcommand{#1}[2][]
{\ifthenelse{\equal{}{##1}}{#2~\eqref{##2}}{#3~\eqref{##1} and~\eqref{##2}}}}

\newcommand{\abstr}[1]{
\begin{abstract}
#1
\end{abstract}}

\newcommand{\bibentry}[8]{

\bibitem[\nospell{#8}]{#1} {\textup #3}. 
\ifthenelse{\equal{}{#6}}
{\newblock \textrm{#4.} \newblock {\em #5}, #7.}
{\newblock \textrm{#4.} \newblock {\em #5, #6}, #7.}
}

\MyMakeTheoMacros{fact}{\fct}{\nfct}

\MyMakeRefMacros{\fctref}{Fact}{Facts}

\MyMakeTheoMacros{observation}{\obs}{\nobs}

\MyMakeRefMacros{\obsref}{Observation}{Observations}

\MyMakeDupTheoMacros{lemma}
{\lem}{\nlem}{\lemdup}{\nlemdup}{\lemapp}{Lemma}{\lemrep}

\MyMakeRefMacros{\lemref}{Lemma}{Lemmas}

\newcommand{\fakelemref}[1]{Lemma~{#1}}

\MyMakeDupTheoMacros{corollary}
{\crl}{\ncrl}{\crldup}{\ncrldup}{\crlapp}{Corollary}{\crlrep}

\MyMakeRefMacros{\crlref}{Corollary}{Corollaries}

\MyMakeTheoMacros{proposition}{\prp}{\nprp}

\newtheorem*{prp*}{\e{Proposition}}

\MyMakeRefMacros{\prpref}{Proposition}{Propositions}

\MyMakeDupTheoMacros{my_claim}
{\clm}{\nclm}{\clmdup}{\nclmdup}{\clmapp}{Claim}{\clmrep}

\MyMakeRefMacros{\clmref}{Claim}{Claims}

\MyMakeDupTheoMacros{theorem}
{\theo}{\ntheo}{\theodup}{\ntheodup}{\theoapp}{Theorem}{\theorep}

\MyMakeRefMacros{\theoref}{Theorem}{Theorems}

\MyMakeTheoMacros{definition}{\defi}{\ndefi}

\MyMakeRefMacros{\defiref}{Definition}{Definitions}

\MyMakeTheoMacros{problem}{\prob}{\nprob}

\MyMakeRefMacros{\probref}{Problem}{Problems}

\MyMakeTheoMacros{protocol}{\prot}{\nprot}

\MyMakeRefMacros{\protref}{Protocol}{Protocols}

\providecommand{\qedsymbol}{\square}

\newcommand{\prf}[2][]{\ifthenelse{\equal{}{#1}}%
{\begin{proof}\renewcommand{\qedsymbol}{$\blacksquare$}%
#2 \end{proof}}%
{\begin{proof}[Proof of #1]%
\renewcommand{\qedsymbol}{$\blacksquare_{\mbox{\it{\scriptsize{#1}}}}$}%
#2 \end{proof}}
}

\newcommand{\prfsk}[2][]{\ifthenelse{\equal{}{#1}}%
{\begin{proof}[Proof sketch]%
\renewcommand{\qedsymbol}{$\blacksquare$} #2 \end{proof}}%
{\begin{proof}[Proof sketch of #1]%
\renewcommand{\qedsymbol}{$\blacksquare_{\mbox{\it{\scriptsize{#1}}}}$}%
#2 \end{proof}}
}

\newcommand{\sect}[2][]{\ifthenelse{\equal{}{#1}}
{\section{#2}}
{\section{#2}\label{#1}}}

\newcommand{\ssect}[2][]{\ifthenelse{\equal{}{#1}}
{\subsection{#2}}
{\subsection{#2}\label{#1}}}

\newcommand{\sssect}[2][]{\ifthenelse{\equal{}{#1}}
{\subsubsection{#2}}
{\subsubsection{#2}\label{#1}}}

\MyMakeRefMacros{\chref}{Chapter}{Chapters}

\MyMakeRefMacros{\sref}{Section}{Sections}

\MyMakeRefMacros{\ssref}{Subsection}{Subsections}

\MyMakeRefMacros{\sssref}{Subsection}{Subsections}

\definecolor{DarkGreen}{rgb}{0,0.45,0.08}
\definecolor{LightBlue}{rgb}{0.122,0.016,0.855}

\newcommand{\IfMathMode}[2]{\ifmmode{#1}\else{#2}\fi}

\newcommand{\Ensuremath}{\ensuremath}

\newcommand{\fbr}[1]{\IfMathMode %
{#1}{$#1$}}                      %

\newcommand{\fnbr}[1]{\mbox{\fbr{#1}}}   %

\newcommand{\fla}[2][*]{\ifthenelse{\equal{}{#1}}{\fbr{#2}}{\fnbr{#2}}}

\newcommand{\mat}[2][]{\ifthenelse{\equal{}{#1}} %
{ \begin{displaymath} #2 \end{displaymath} } %
{ \begin{equation} \label{#1} #2 \end{equation} }%
}

\newcommand{\matal}[2][]{\mat[#1]{\begin{aligned} #2 \end{aligned}}}

\newcommand{\f}{\fla}

\newcommand{\m}{\mat}
\newcommand{\mal}{\matal}

\newcommand{\twocase}[4]%
{\begin{cases}#1 &\txt{#2}\\#3 &\txt{#4}\end{cases}}

\MyMakeEqRefMacros{\equref}{Equation}{Equations}

\MyMakeEqRefMacros{\expref}{Expression}{Expressions}

\MyMakeEqRefMacros{\inequref}{Inequality}{Inequalities}

\newcommand{\bracref}[1]{(\ref{#1})}

\newcommand{\bref}{\bracref}

\MyMakeRefMacros{\figref}{Figure}{Figures}

\providecommand{\middle}{\big}

\newmatop{\tr}{tr}

\newcommand{\KL}[2]{d_{KL}\llp{#1}\middle|\middle|{#2}\rrp}

\providecommand{\E}[2][]{\mathop{\mathbf{E}}_{#1}{\left[{#2}\right]}}

\newcommand{\PR}[2][]{\mathop{\mathbf{Pr}}_{#1}{\left[{#2}\right]}}
\newcommand{\PRr}[3][]{\mathop{\mathbf{Pr}}_{#1}{\left[{#2}\middle|\vphantom{|_1^1}{#3}\right]}}

\newcommand{\U}[1][]{\ifthenelse{\equal{}{#1}}%
{{\cal U}}%
{{\cal U}_{#1}}}

\newcommand{\wt}{\widetilde}

\newcommand{\pl}[1][]{\nospell{\ifthenelse{\equal{}{#1}}%
{\!\stackrel'{}\!\!\txt{s}}%
{\fla{#1\!\stackrel'{}\!\!\txt{s}}}}}

\newcommand{\ord}[1][]{\nospell{\ifthenelse{\equal{}{#1}}%
{\txt{'th}}%
{\ifthenelse{\equal{1}{#1}}{$1\txt{'st}$}{\ifthenelse{\equal{2}{#1}}{$2\txt{'nd}$}{\ifthenelse{\equal{3}{#1}}{$3\txt{'rd}$}{\fla{#1\txt{'th}}}}}}}}

\newmat[]{\llp}{\left(}
\newmat[]{\rrp}{\right)}

\newmat[]{\lla}{\left\langle}
\newmat[]{\rra}{\right\rangle}

\newmat[]{\llf}{\left\lfloor}
\newmat[]{\rrf}{\right\rfloor}

\newmat[]{\dt}{\cdot}
\newmat[]{\tm}{\cdot}
\newmat[]{\xor}{\oplus}
\newmat[]{\sbseq}{\subseteq}
\newmat[]{\sbs}{\subset}
\newmat[]{\smin}{\setminus}
\newmat[]{\eps}{\varepsilon}
\newmat[]{\deq}{\stackrel{\textrm{def}}{=}}

\providemat{\QQ}{\mathbb{Q}}
\providematarg{\NN}{\ifthenelse{\equal{}{#1}}%
{\mathbb{N}}%
{\mathbb{N}_{#1}}}
\providematarg{\CC}{\ifthenelse{\equal{}{#1}}%
{\mathbb{C}}%
{\mathbb{C}^{#1}}}

\newmat[]{\ds}{\dots}
\newmat[]{\dc}{,\ds,}

\newmat[]{\sdle}{\preceq}

\newcommand{\enum}[1]{\begin{enumerate} #1 \end{enumerate}}

\newcommand{\itemi}[2][]{\ifthenelse{\equal{}{#1}}
{\begin{itemize} #2 \end{itemize}}
{\begin{itemize}[#1] #2 \end{itemize}}}

\newcommand{\wrt}{w.r.t.\ }	 %

\newcommand{\wlg}{w.l.g.\ }	 %

\newcommand{\eg}{e.g., }	 %

\newcommand{\fr}[3][*]{%
\ifthenelse{\equal{*}{#1}}        %
{\frac{#2}{#3}}{}%
\ifthenelse{\equal{/}{#1}}        %
{\nicefrac{#2}{#3}}{}%
\ifthenelse{\equal{}{#1}}         %
{\left.#2\middle/#3\right.}{}%
\ifthenelse{\equal{p_}{#1}}       %
{\left.\left(#2\right)\middle/#3\right.}{}%
\ifthenelse{\equal{_p}{#1}}       %
{\left.#2\middle/\left(#3\right)\right.}{}%
\ifthenelse{\equal{pp}{#1}}       %
{\left.\left(#2\right)\middle/\left(#3\right)\right.}{}
}

\newcommand{\dr}{\nicefrac}

\newcommand{\sq}{\mathpalette\MySQRT}  %

\def\MySQRT#1#2{    %
\setbox0=\hbox{$#1\sqrt{#2\,}$}\dimen0=\ht0%
\advance\dimen0-0.2\ht0%
\setbox2=\hbox{\vrule height\ht0 depth -\dimen0}%
{\box0\lower0.4pt\box2}}

\newcommand{\set}[2][]{\ifthenelse{\equal{}{#1}} %
{\Ensuremath{\left\{#2\right\}}}%
{\Ensuremath{\left\{#2\,\middle\arrowvert\,#1\right\}}}}

\newcommand{\Minn}[3][]{\ifthenelse{\equal{}{#1}} %
{\Ensuremath{\min_{#2}{\left\{#3\right\}}}}%
{\Ensuremath{\min_{#2}{\left\{#3\,\middle\arrowvert\,#1\right\}}}}}

\newcommand{\Max}[2][]{\ifthenelse{\equal{}{#1}} %
{\Ensuremath{\max{\left\{#2\right\}}}}%
{\Ensuremath{\max{\left\{#2\,\middle\arrowvert\,#1\right\}}}}}

\newcommand{\Maxx}[3][]{\ifthenelse{\equal{}{#1}} %
{\Ensuremath{\max_{#2}{\left\{#3\right\}}}}%
{\Ensuremath{\max_{#2}{\left\{#3\,\middle\arrowvert\,#1\right\}}}}}

\newcommand{\newfunction}[2]{ %
\newcommand{#1}[2][*]{\ifthenelse{\equal{*}{##1}}%
{\Ensuremath{#2{\left(##2\right)}}}%
{#2(##2)}}%
}

\newfunction{\asO}{O}
\newfunction{\asOm}{\Omega}
\newfunction{\asT}{\Theta}

\providecommand{\ket}[1]{\Ensuremath{\left|#1\rra}}

\newcommand{\kbra}[2][]{\ifthenelse{\equal{}{#1}}%
{\Ensuremath{\left|#2\middle\rangle\hspace{-2.5pt}\middle\langle #2\right|}}%
{\Ensuremath{\left|#1\middle\rangle\hspace{-2.5pt}\middle\langle #2\right|}}%
}
\newcommand{\bket}[3][]{\ifthenelse{\equal{}{#1}}%
{\Ensuremath{\lla #2\middle|#3\rra}}%
{\Ensuremath{\lla #2\middle|#1\middle|#3\rra}}
}

\mydef{\01}{\set{0,1}}

\renewcommand{\l}{\left}
\renewcommand{\r}{\right}

\newcommand{\sz}[2][]{\ifthenelse{\equal{}{#1}}%
{\Ensuremath{\left|#2\right|}}%
{\Ensuremath{\left|#2\middle|_{#1}\right.}}}

\providecommand{\floor}[2][*]{\ifthenelse{\equal{}{#1}}%
{\lfloor #2 \rfloor}%
{\llf #2 \rrf}}

\newcommand{\fn}{\footnote}

\newcommand{\nin}{\not\in}   %

\newcommand{\e}{\emph}

\newcommand{\bm}{\mathbf}  %

\newcommand{\ul}[1]{\underline{#1}}  %

\newcommand{\txt}[1]{\textrm{#1}}   %

\newcommand{\Cl}[1]{{\cal #1}} %

\newcommand{\tb}{\quad}
\newcommand{\tbb}{\qquad}

\DeclareMathAlphabet{\lowcal}{OT1}{pzc}{m}{it}

\usepackage{hyphenat}   %

\date{}
\newident{\hmp}{HMP}
\newident{\hf}{HMP_4}

\newmat[]{\GH}{\Cl G_{\hf}}
\newmat{\sH}{\sigma_{HM}}

\newmat{\Ch}{\Cl C}
\newmat{\cn}{\bm c}
\newmat{\x}{\bm x}

\newmat[]{\Lh}{L_{\lowcal{hl}}}
\newmat[]{\Lb}{L_{\lowcal{bn}}}

\newident{\Ver}{\lowcal{Ver}}

\newident{\qm}{\mathcal Q}

\newidentarg{\ka}{\ket{\alpha(#1)}}
\newidentarg{\kba}{\kbra{\alpha(#1)}}

\title{Quantum Money with Classical Verification}

\author{
{\bf Dmitry Gavinsky}\\
{\small NEC Laboratories America, Inc.}\\
{\small Princeton, NJ, U.S.A.}
}

\begin{document}

\maketitle

\thispagestyle{empty}

\abstr{
We propose and construct a quantum money scheme that allows verification through \e{classical} communication with a bank.
This is the first demonstration that a secure quantum money scheme exists that does not require quantum communication for coin verification.

Our scheme is secure against adaptive adversaries -- this property is not directly related to the possibility of classical verification, nevertheless none of the earlier quantum money constructions is known to possess it.

}

\sect{Introduction}

In 1983 Wiesner~\cite{W83_Co_Co} proposed a new quantum cryptographic scheme, that later became known as \e{quantum money}.
Informally, a \e{quantum coin} is a unique object that can be created by a trusted \e{bank}, then circulated among untrusted \e{holders}.\fn
{The notation is still unset in this relatively new area of research.
In particular, each coin in our construction will have its own identification number, and some authors would call such items \e{quantum banknotes}, to emphasize the uniqueness.
Also, what we call a \e{bank} is sometimes addressed as a \e{mint}.}
A holder of a coin should be able to verify it, and the verification must confirm that the coin is authentic if it has been circulated according to the prescribed rules.
On the other hand, if a holder wants to counterfeit a coin, that is, to create several objects such that each of them would pass verification, he must fail in doing so with overwhelmingly high probability.

Wiesner has demonstrated that quantum mechanics (as opposed to classical physics) allows money schemes, and the basic principle that made such constructions possible was that of \e{quantum uncertainty}.
The principle states that there are properties of a quantum object that are known to its ``manufacturer'' but cannot be learnt by an observer who measures the object; nevertheless, those properties can be later ``verified'' by the manufacturer.
Accordingly, a bank can prepare objects with this kind of ``secret properties'' and let the holders use them as quantum coins -- not knowing the secrets, untrusted holders would not be able to forge counterfeits.

\ssect{Prior work}

In Wiesner's original construction~\cite{W83_Co_Co,BBBW83_Qu_Cr} a coin had to be sent back to the bank in order to get verified.
This could be viewed as a possible drawback:\ a coin might get ``stolen'', or intentionally ``ruined'' by an adversary who had access to the communication channel between a coin holder and the bank.

This problem has been addressed in a number of works.
The approach taken by Aaronson~\cite{A09_Qu_Co}, Lutomirski et al.~\cite{LAFGK10_Bre}, Farhi et al.~\cite{FGHLS10_Qu_Mo} and in the upcoming Aaronson and Christiano~\cite{AC12_Qu_Mo} was to allow the holders to verify quantum coins locally, not having to contact the bank.
Clearly, in this situation an adversary can, given unlimited computational resources, produce as many counterfeit coins as he wishes (being able to locally verify implies having a complete description of all objects that would pass the verification, so coin forgery becomes an achievable, albeit possibly computationally-expensive task).
What is worse, the present state of mathematical development only allows to \e{conjecture} that certain tasks are hard for a reasonably powerful model of computation, and a major breakthrough would be required to argue that a scheme of this type is secure, say, against an adversary who can use a Turing machine.

In a different line of research, Tokunaga, Okamoto and Imoto~\cite{TOI03_An_Qu} and Mosca and Stebila~\cite{MS10_Qu_Co} considered the problem of creating quantum money that can be used \e{anonymously}.\fn
{Note that locally verifiable coins can be viewed as a partial answer to this requirement:\ when the bank isn't involved in the verification procedure it cannot ``trace'' the transactions.}
In~\cite{TOI03_An_Qu} a coin holder introduces some local randomness into the state of a coin to obtain anonymity.
In~\cite{MS10_Qu_Co} the construction allows multiple identical (but still resistant to counterfeiting) instances of quantum coins.
In both of these works quantum communication with a bank is required in order to use the scheme (\cite{MS10_Qu_Co} discusses the hypothetical possibility of using computational hardness assumptions to allow local verification).

Relatively recently another limitation of all previously known quantum money schemes has been noticed by Aaronson~\cite{A09_Qu_Co} and by Lutomirski~\cite{L10_An_O}:
An adversary can gain even more power from interacting \e{adaptively} with the bank.
No quantum money scheme was know to be resistant to this type of attacks; in fact, \cite{L10_An_O} has shown a very efficient adaptive attack against one version of Wiesner's scheme (which was unconditionally secure against non-adaptive adversaries).

\ssect[ss_our]{Our results}

In this work we propose to use \e{classical communication with a bank} in order to verify a quantum coin.
We construct such a scheme.
This is the first demonstration that \e{a secure quantum money scheme exists that does not require quantum communication for coin verification}.

Some advantages of our construction over the previously known ones are:\itemi{
\item Unlike the original scheme of Wiesner and the constructions of Mosca and Stebila, \e{our construction does not require quantum communication with a bank} in order to verify a coin.
\item We \e{prove} that our scheme is (unconditionally) secure; security arguments for schemes with local verification require either unproved hardness assumptions or a major mathematical breakthrough (complexity lower bounds).
Moreover, to the best of our knowledge, no such scheme has been shown to be secure under so-called ``widely believed'' unproved assumptions.\fn
{It has happened that a proposed scheme was broken soon after its publication.}
\item Unlike the schemes with local verification, our construction remains \e{secure against computationally unlimited adversary} who obeys the laws of quantum mechanics.
}

Besides offering possible practical advantages, the concept of quantum money with classical verification gives rise to natural and attractive theoretical questions.

Another advantage of our construction is not directly related to the possibility of quantum verification:\itemi{
\item Our scheme remains secure against an adversary who uses \e{adaptive multi-round attacks}; no such scheme was known before.
} 

Note that adaptive multi-round attacks are also conceivable in the case of money schemes with quantum verification alone:\ an adversary can, for example, ``split'' a coin into two ``fragments'', send one of them to the bank and collect the response, and later use the remaining fragment in a way that would depend on the bank's response to the first fragment.
Indeed, Lutomirski~\cite{L10_An_O} has demonstrated a linear-time adaptive attack against one version of Wiesner's scheme, which was provably secure against non-adaptive adversaries.
Before this work it was open whether any quantum money scheme can be resistant to adaptive multi-round attacks.

We call our quantum money scheme \qm.
In order to verify a \qm-coin a holder has to contact the bank via a classical communication channel and perform quantum measurements, as directed by the bank, then report the outcomes.
In the end the bank either confirms that the coin is valid or rejects it.

Our construction has the following specific properties:
\itemi{
\item The coins are \e{exponentially} hard to counterfeit (cf.\ \theoref{t_main} and \crlref{c_main}).
\item The classical communication channel used for verification can be \e{unencrypted}:\ e.g., both the bank and the coin holder can broadcast their messages, without compromising security of the scheme.
\item Our scheme remains secure against an adversary who uses \e{adaptive ``attempted verifications''} in order to collect information about a coin.
Exponentially many such attempts have to be made before one has non-negligible chances to counterfeit a coin.
\item The database of the bank is \e{static}, and therefore many de-centralized ``verification branches'' can exist that do not have to communicate with one another.
\item The number of verifications that a \qm-coin can go through is limited -- the number of qubits required to store a coin is polynomial in the number of validity tests via classical communication that the coin can go through during its circulation period (after that it would have to be replaced by the bank).
We show that this dependency is \e{optimal} (cf.\ \theoref{t_opti}).
}

\ssect{Related work}

Using a different approach, Aaronson and Christiano in the upcoming~\cite{AC12_Qu_Mo} will construct a scheme that uses quantum communication with a bank for verification (like Wiesner's original scheme) and is resistant against adaptive multi-round attacks.

Very recently some of the ideas proposed in this work have been further developed by Pastawski et al.~\cite{PYJLC11_Un_No} and by Molina et al.~\cite{MVW12_Op_Co}.

\sect{Who needs quantum money?}

The first quantum money scheme was proposed by Wiesner more than 30 years ago (several years before~\cite{W83_Co_Co} was published).
Nevertheless, there seems to remain some confusion about the advantages that quantum money has over possible classical constructions.
Below we reproduce a typical ``classical money'' proposal, then discuss the advantages of Wiesner's scheme, then further advantages of our construction.

Note that here we are not comparing our scheme to the previously known ones (that was the subject of \sref{ss_our}).
Instead, this part (informally) addresses the question posed by its title.

\ssect{A classical proposal}

Let every coin issued by the bank contain a secret string $s$, known only to the bank and to the current coin holder.
When a coin holder Alice wants to pass her coin to a new coin holder Bob, they run the following protocol:\itemi{
\item Alice sends to the bank the string $s$ and tells the bank that she wants to pass the coin to Bob.
\item The bank checks that $s$ is a valid secret string (if not then a forgery attempt has been detected), then erases $s$ from the list of valid strings and adds to the list a newly generated secret string $s'$.
\item The bank sends $s'$ to Bob; henceforth, Bob holds the coin.
}

\ssect{Advantages of Wiesner's scheme}

\itemi{
\item The bank's database can be static (for the classical scheme to be secure, it is crucial that a new secret string is issued each time a coin is passed along).
\item Interaction with the bank does not require \f 3-party authentication (for the classical scheme to be secure, the bank has to make sure that the \e{only} recipient of the newly generated secret string is the party named by Alice in the first round).
}

\ssect{Advantages of our scheme}

\itemi{
\item All the benefits of Wiesner's construction listed above.
\item The communication channel can be classical and not encrypted.
Moreover, all the messages (both ways) can be openly broadcast.
\item In the classical scheme, as well as in Wiesner's scheme, an intruder who pretends to be the bank can steal a valid coin from its fair holder who wants to verify it.
Our scheme shields against that.
}

\sect{Notation and preliminaries}

For $a\in\NN$ we denote $[a]\deq\set{1\dc a}$.
Denote by $I_a$ the identity matrix of rank $a$.
For any finite $A$ we denote by $\U[A]$ the uniform distribution over the elements of $A$.

We will use concentration bounds extensively in our proofs.

\ntheo[t_Che]{(Chernoff bound)}{Let $X_1\dc X_n$ be mutually independent random variables taking values in $[0,1]$, such that $\E{X_i}=\mu$ for all $i\in[n]$.
Then for any $\lambda>0$,
\m{\PR{\sum_{i\in[n]}X_i\ge(1+\lambda)\mu n}\le
e^{\fr{-n\lambda^2\mu}{2+\lambda}},}
and
\m{\PR{\sum_{i\in[n]}X_i\le(1-\lambda)\mu n}\le
e^{\fr{-n\lambda^2\mu}2}.}
}

We also need a generalization, originally proved by Panconesi and Srinivasan~\cite{PS97_Ran}.
The following version of it is due to Impagliazzo and Kabanets~\cite{IK10_Con}.

\ntheo[t_GenChe]{(Generalized Chernoff bound)}{Let $X_1\dc X_n$ be Boolean random variables, such that for some $\delta$ and every $S\sbseq[n]$ it holds that $\PR{\wedge_{i\in S}X_i=1}\le\delta^{|S|}$.
Then
\m{\PR{\sum_{i\in[n]}X_i\ge(1+\lambda)\delta n}\le
e^{-2n\lambda^2\delta^2}.}
}

\crl[c_Che+]{Let $X_1\dc X_n$ be Boolean random variables, such that for all $i\in[n]$ and any event $C$ that only depends on $\{X_j|j\neq i\}$ it holds that $\PRr{X_i=1}C\le\delta$.
Then
\m{\PR{\sum_{i\in[n]}X_i\ge(1+\lambda)\delta n}\le
e^{-2n\lambda^2\delta^2}.}
}

\prf{For every $S\sbseq[n]$,
\m{\PR{\wedge_{i\in[|S|]}X_{S_i}=1}=
\prod_{i\in[|S|]}\PRr{X_{S_i}=1}{X_{S_1}=\ds=X_{S_{i-1}}=1}
\le\delta^{|S|},}
where $S_i$ is the \ord[i] least element of $S$.
}

We will also need the following combinatorial lemma (a rather standard one, \eg see \fakelemref{2.2} in Jukna~\cite{J01_Ex}).

\lem[l_sets]{Let $A_1\dc A_N$ be subsets of $[n]$ of average size $t$.
Suppose that $\sz{A_i\cap A_j}\le s$ for every $i\neq j$.
Then either $N<\fr[/]{2n}t$ or $s>\fr[/]{t^2}{2n}$ (or both).\fn
{The asymptotic guarantees of our \lemref{l_sets} are slightly better than those of \fakelemref{2.2} in~\cite{J01_Ex} -- there the main statement is more general, but the result is weaker in the special case that we are interested in.}}

\prf{For $x\in[n]$, let $d(x)\deq\sz{\set[x\in A_i]{i}}$.
On the one hand,
\m{\sum_{x=1}^nd(x)=\sum_{i=1}^N\sz{A_i}=Nt
\tb\implies\tb
\sum_{x=1}^n d(x)^2\ge\fr{N^2t^2}n.}
On the other hand,
\m{\sum_{x=1}^n\l(d(x)\r)^2
=\sum_{i=1}^N\sum_{j=1}^N\sz{A_i\cap A_j}
=\sum_{i=1}^N\sz{A_i}
+\sum_{j\neq i}\sz{A_i\cap A_j}
\le Nt+N(N-1)s.}
Therefore,
\m{\fr{Nt^2}n<t+Ns
\tb\implies\tb
s>\fr{t^2}n-\fr{t}N,}
and the result follows.}

\sect{Our quantum money scheme \qm}

One of the main technical ingredients of our construction is a constant-dimensional ($n=4$) special case of a relational communication problem called \e{Hidden Matching Problem (\hmp)}, first considered by Bar-Yossef, Jayram and Kerenidis~\cite{BJK04_Expo} in the context of communication complexity.

\defi{Let \hf\ be as follows.
For $x\in\01^4$ and $m,a,b\in\01$, we say that $(x,m,a,b)\in\hf$ if $b=\twocase{x_1\xor x_{2+m}}{if $a=0$}{x_{3-m}\xor x_4}{if $a=1$}$.}

Intuitively, if we view $x\in\01^4$ as a binary coloring of $4$ vertices then a tuple $(x,m,a,b)$ satisfies the relation \hf\ if and only if $b$ indicates whether $x$ assigns distinct colors to the pair of vertices determined by $m$ and $a$.

It has been shown in~\cite{BJK04_Expo} that if Alice receives $x$ and Bob receives $m$ then Alice can send a short \e{quantum} message to Bob that would allow him to produce a valid answer $(a,b)$; on the other hand, if Alice is only allowed to send classical bits then a much longer message is required.
The authors were interested in the asymptotic behavior of quantum and classical communication cost of \hmp, and they gave an elegant proof that the gap between the two is exponential.

How can it help us?
We want to build a scheme that would be safe against both classical and quantum attacks; moreover, we want to be able to carry out certain communication task (testing validity of a coin) using \e{only classical} communication.
So, why are we interested in something showing that quantum communication is more powerful than classical?

The answer is that the role of quantum communication from~\cite{BJK04_Expo} in our case is played by a \e{quantum coin}:\ when the bank issues a coin, it sends a quantum message to its future holder.
The core of our construction is the observation (apparently, new to this work) that in certain quantum one-way protocol for \hmp, a single message from Alice cannot be used by Bob in order to produce valid answers \wrt several different values of $m$.
In other words, \e{the message cannot be ``reused''}.
This holds in spite of the fact that a message from Alice cannot depend on $m$, thus using it Bob can produce a valid answer \wrt any legitimate value of $m$.

In our construction we will use a state \ka[x] of $2$ qubits (corresponding to the quantum message that Alice would send to Bob in a one-way protocol for \hf) that allows its holder, who is given $m$ but doesn't know $x$, to find an ``answer'' $(a,b)$ that satisfies \hf\ with certainty.
On the other hand, using the same state in order to find $(a_0,b_0)$ and $(a_1,b_1)$, such that $(x,m,a_m,b_m)\in\hf$ for both $m=0$ and $m=1$ would fail with probability at least $\dr14$.
In other words, our state of $2$ qubits will be \e{useful but not reusable} for producing an answer to \hf.

Let the bank choose $x_1\dc x_k\in\01^4$ at random, keep them in secret and produce quantum states $\ka[x_1]\dc\ka[x_k]$.
A newly issued \qm-coin consists of a piece of paper glued to $k$ quantum registers that hold $\ka[x_1]\dc\ka[x_k]$.
The piece of paper contains a unique identification tag and $k$ initially unmarked positions, where the \ord[i] position has to be marked when the corresponding \ka[x_i] is used in the verification protocol.

More formally:
\ndefi{(\hf-states)}{Let $x\in\01^4$.
The corresponding \hf-state is 
\m{\ka[x]\deq\fr12\tm\sum_{1\le i\le4}(-1)^{x_i}\ket i.}}

Interestingly, the \hf-states (in their multidimensional version) were first considered by Kerenidis and de Wolf~\cite{KW04_Ex_Lo} in order to prove a lower bound on the length of certain codes, and that was before the Hidden Matching Problem was defined.

\ndefi{(\hf-queries)}{An \hf-query is an element $m\in\01$.
A valid answer to the query \wrt $x\in\01^4$ is a pair $(a,b)\in\01\times\01$, such that $(x,m,a,b)\in\hf$.}

An \hf-state can be used to answer an \hf-query with certainty:
If $m=0$, let
\m{v_1\deq\fr{\ket1+\ket2}{\sq2}, v_2\deq\fr{\ket1-\ket2}{\sq2},
v_3\deq\fr{\ket3+\ket4}{\sq2}, v_4\deq\fr{\ket3-\ket4}{\sq2};}
otherwise ($m=1$), let
\m{v_1\deq\fr{\ket1+\ket3}{\sq2}, v_2\deq\fr{\ket1-\ket3}{\sq2},
v_3\deq\fr{\ket2+\ket4}{\sq2}, v_4\deq\fr{\ket2-\ket4}{\sq2}.}
Measure \ka[x] in the basis \set{v_1,v_2,v_3,v_4}, and let $(a,b)$ be $(0,0)$ if the outcome is $v_1$; $(0,1)$ in the case of $v_2$; $(1,0)$ in the case of $v_3$; $(1,1)$ in the case of $v_4$.
Then $(x,m,a,b)\in\hf$ always.

\ndefi{(\qm-coins)}{Let $3|t$.
A secret record consists of $k$ entries $x_1\dc x_k$, $x_i\in\01^4$ (i.e., the secret record contains $4k$ bits).

A ``fresh'' \qm-coin corresponding to the record $\l(x_1\dc x_k\r)$ consists of\itemi{
\item $k$ quantum registers consisting of $2$ qubits each, where the \ord[i] register contains \ka[x_i];
\item a $k$-bit classical register $P$, that is initially set to $0^k$;
\item a unique identification number.
}
}

A bank produces \e{fresh} \qm-coins; as a \qm-coin goes through more and more verification protocols, its quantum registers lose their original content (and that shall be reflected in the corresponding bits of $P$, see below).
The identification number of every coin issued by the bank must be unique.

To verify a \qm-coin through classical communication with the bank, its holder runs the following protocol \Ver\ ($t$ is a parameter in the construction of \qm\ that will be polynomially related to $k$).

\prot{When a holder of a valid \qm-coin follows the protocol, verification goes like this:}
\enum{
\item \label{step_ini} The holder sends the identification number of the \qm-coin to the bank.
\item \label{step_t} The bank chooses uniformly at random a set $\Lb\sbs[k]$ of size $t$, and sends it to the coin holder.
\item \label{step_2t/3} The holder consults with $P$ and chooses uniformly at random a set $\Lh\sbs\Lb$ consisting of $\dr{2t}3$ yet unmarked positions.
He sends $\Lh$ to the bank and marks in $P$ all the elements of $\Lh$ as used.
\item \label{step_m} The bank chooses at random $\dr{2t}3$ values $m_i\in\01$, one for each $i\in \Lh$, and sends them to the coin holder.
\item \label{step_meas} The holder measures the quantum registers corresponding to the elements of $\Lh$ in order to produce $\dr{2t}3$ pairs $(a_i,b_i)$, such that $(x_i,m_i,a_i,b_i)\in\hf$ for all $i\in \Lh$.
He sends the list of \pl[(a_i,b_i)] to the bank.
\item \label{step_verdict} The bank checks whether $(x_i,m_i,a_i,b_i)\in\hf$ for all $i\in \Lh$, in which case it confirms validity of the \qm-coin.
Otherwise, the coin is declared to be a counterfeit.
}

We will say that an instance of \Ver\ has been \e{passed} or \e{won} if the bank's final response was ``valid''.

Observe that a fair coin holder fails to pass \Ver\ with exponentially small probability (corresponding to the situation when less than $\dr k4$ of the coin registers are marked as used, but among the $t$ registers that were uniformly chosen by the bank more than $\dr t3$ are marked as used).
If this happens, a new run of \Ver\ can be started.

It follows from the earlier discussion that both the bank and a fair coin holder can perform their parts of \Ver\ efficiently.
Note also that the secret records kept by the bank do not change as a result of executing \Ver\ -- that is, the bank's database is \e{static}.

Intuitively, adversarial ability to counterfeit a \qm-coin shall imply ability to answer \wrt the same quantum register $i$ both to the question $m_i=0$ and to $m_i=1$.
As we said before, that can be done with probability at most $\dr34$; moreover, it turns out that in order to successfully counterfeit a coin the adversary must be able to answer both the \hf-queries \wrt a considerable fraction of the coin's registers, and that will imply exponentially small probability of adversarial success.
We will formalize and prove this intuition in \sref{s_sect}.

We will show (cf.\ \theoref{t_main} and \crlref{c_main}) that only after an adversary has run $e^{\asOm{\dr{t^3}{k^2}}}$ auxiliary instances of \Ver, he might be able to counterfeit a \qm-coin with success probability higher than $e^{-\asOm{\dr{t^2}k}}$.

Note that every run of \Ver\ ``costs'' $\dr{2t}3$ yet unused quantum registers.
As soon as $\dr k4$ registers have been used, the \qm-coin has to be returned to bank (the bank still would be able to verify its validity and issue a replacement).
Accordingly, after $\floor{\dr{3k}{8t}}$ runs of \Ver\ a \qm-coin has to be returned to the bank.

\ul{\e{To conclude}}:
Choosing, for example, $t\in\asT{k^{\dr34}}$ gives a construction where a coin that consists of $2k$ qubits can go through \asOm{k^{\dr14}} validity tests via classical communication with the bank, and where it takes $e^{k^{\asOm1}}$ time to forge a counterfeit with probability higher than $e^{-k^{\asOm1}}$.
The bank's secret database contains $4k$ bits corresponding to every coin, and those records are static (in particular, many de-centralized ``verification branches'' can exist that do not have to communicate with one another).
In \sref{s_opti} we will show that these parameters are very close to the best possible.

\sect[s_sect]{Security of \qm}

We are giving ``extended security guarantees'', as follows.
Instead of only arguing that the first cheating attempt is not likely to succeed, we allow an adversary to use \e{multiple attacking attempts} -- namely, even having been caught cheating in the past, he may continue his attempts.
Recall that we allow adaptive attacks, thus something learnt form the earlier attempts might help the adversary in future attacks.

Informally, our security guarantees will be expressed like this:
\e{An exponentially large number of partially completed instances\fn
{By a partially completed protocol we mean an instance of \Ver, where the first response from the bank has been received and analyzed by the adversary.}
of \Ver\ are required for an adversary to have non-negligible probability to make a counterfeit coin}.

A high-level view of our security analysis is as follows.
First we make preliminary observations regarding possible attacks on the \qm-scheme (\sref{ss_pos_at}), and demonstrate useful properties of \hf-states (\sref{ss_hmp}).
Then we claim that counterfeiting a \qm-coin has its ``cost'', in terms of the number of preliminary runs of \Ver\ that are required in order to collect enough auxiliary information about the coin (\sref{ss_cost}).
Finally, we reduce unrestricted attacks to more structured ones and show their limitations (\sref[ss_gen_to_near-phased]{ss_phased_slow}, respectively).
We conclude in \sref{ss_together} that exponentially many preliminary runs of \Ver\ are required in order to counterfeit a \qm-coin.

\ssect[ss_pos_at]{Possible attacks and security guarantees}

Our goal will be to show that a \qm-coin is hard to counterfeit.
First, we want to argue that in order to establish security of our \qm-scheme it is enough to consider the situation when starting with a fresh authentic \qm-coin, an adversary runs many instances of \Ver\ (probably, in a non-consecutive manner) for this coin\fn
{Note that every run of \Ver\ can be associated with certain \qm-coin via its identification number, as reported by the coin holder in the first round of a protocol.},
and his goal is to produce two (possibly, entangled) quantum objects that have non-negligible probability to be accepted by the bank as valid coins.

Probably, the most harmful attack on \qm\ would be the one where an adversary starts with $M$ fresh \qm-coins, and his goal is to produce $M+1$ quantum objects that are all likely to be accepted as valid \qm-coins.\fn
{Several modifications of this cheating setup can be considered, but it seems that all of them can be reduced to the ``$M+1$ out of $M$'' regime.}
Let us look at the ``two out of one'' security guarantee that we give for the \qm-scheme, and see how it implies robustness against ``multi-coin'' attacks.

Let us call \e{the first response} a message that the bank sends in step $2$ of \Ver\ (that is, a list of $t$  positions).
In \sref{ss_together} we establish the following theorem.

\theodup{t_main}{\teoMain}{Let a fresh \qm-coin be given to a computationally unlimited adversary who runs auxiliary instances of \Ver\ for this coin and produces two (possibly, entangled) ``counterfeits'' $\rho_1$ and $\rho_2$.
Then
\m{U\in e^{\asOm{\fr[/]{t^3}{k^2}}}}
exists, such that if the adversary has received and analyzed the first bank's responses in at most $U$ instances of \Ver, then the probability that both $\rho_1$ and $\rho_2$ pass \Ver\ is in 
\m{e^{-\asOm{\fr[/]{t^2}k}}.}}

\crl[c_main]{Let $M$ fresh \qm-coins be given to a computationally unlimited adversary who analyzes the first bank's responses in at most $U$ auxiliary instances of \Ver, for $U$ as in \theoref{t_main}.
If the adversary outputs $M+1$ quantum objects then the probability that all of them pass \Ver\ is in $e^{\ln M-\asOm{\fr[/]{t^2}k}}$.}

\prf{If the identification numbers of the $M+1$ produced quantum objects are not a subset of the identification numbers of $M$ initially given objects then at least one counterfeit has been produced ``from scratch'', and it is easy to see that the probability of success in this case is negligible.

Otherwise there is at least one identification number that appears more than once among the $M+1$ produces quantum objects with probability at least $\dr1M$.
Starting with a single coin, one can emulate the cheating strategy for $M$ coins by locally creating $M-1$ \qm-coins and running the protocol, locally computing bank's responses according to \Ver\ \wrt any of the $M-1$ auxiliary coins.
If in the end of emulation at least two object are marked with the same identification number as the given coin then those two objects are returned, otherwise arbitrary output is produced.

If the \f M-coin counterfeit strategy produces $M+1$ quantum objects that successfully pass verification with probability $\eps$, then the strategy above succeeds in counterfeiting a single coin with probability at least $\dr\eps M$, and the corollary follows from \theoref{t_main}.}

\ssect[ss_hmp]{Quantum retrieval games}

To analyze some useful properties of \hf-states we define the notions of \e{quantum retrieval games}, \e{physical projections}, and \e{selective projections}.

\ndefi{(quantum retrieval games)}{Let $k,m,n\in\NN$, $\sigma\sbseq[m]\times[n]$, and $\forall a\in[n]$ let $\rho_a\in\CC^{k\times k}$ be positive semidefinite such that $\tr\llp\sum_a\rho_a\rrp=1$.
Then $\Cl G=\llp(\rho_a)_{a\in[n]},\sigma\rrp$ is a quantum retrieval game.}

The notion of quantum retrieval games is aimed to model the situation when a mixed quantum state $\sum\rho_a$ is measured in order to extract some information about $a$.
The relation $\sigma$ describes what knowledge is wanted.
We will consider situations when an \f m-outcome quantum measurement is applied to $\sum\rho_a$, and we say that the game $\Cl G$ has been won if the pair $\llp\langle outcome~of~the~measurement\rangle,a\rrp$ is in $\sigma$.
Formally:

\ndefi{(selective and physical projections)}{Let $\Cl P=\set{P_i}_{i=1}^m$ be a set of projections in $\CC^{k\times k}$, s.t.\ $\sum_i P_i\sdle I$.
Call $\Cl P$ a selective projection.
A selective projection is called \e{physical projections} if it satisfies $\sum_i P_i=I$.}

\ndefi[def_values]{(selective and physical values of a game)}{The value of $\Cl G$ \wrt $\Cl P$ is defined as
\m{\fr{\sum_{(i,a)\in\sigma}\tr(P_i\rho_a)}{\sum_{i,a}\tr(P_i\rho_a)},}
and if $\Cl P$ is a selective projection then the value is undefined unless $\sum_{i,a}\tr(P_i\rho_a)>0$.
The selective value of $\Cl G$ is the supremum of the game's value \wrt selective projections, and the physical value of $\Cl G$ is the supremum of the game's value \wrt physical projections.}

Note that for physical projections it holds that $\sum_{i,a}\tr(P_i\rho_a)=1$ (and the above definition simplifies to $\sum_{(i,a)\in\sigma}\tr(P_i\rho_a)$ in that case).
Physical projections are the most general ``mechanism'' offered by quantum mechanics to extract classical information from a quantum state.

Selective projections are, in general, more powerful than physical projections (they correspond to measurements with ``postselection'', and those are not allowed by the laws of quantum mechanics).
We will consider selective projections in some of our impossibility statements, that will allow simpler proofs of direct product statements that we will need.
Like in the case of physical projections, we will view the elements of $\Cl P$ as outcomes.
Clearly, the selective value of a game is always at least as large as its physical value.

A physical projection $\Cl P$ corresponds to some POVM measurement, and the elements of $\Cl P$ are the possible outcomes.
When it is applied to some (normalized) $\rho\in\CC^{k\times k}$, the \ord[i] outcome occurs with probability $\tr(P_i\rho)$.
If \ord[i] outcome occurred then the state of the quantum register that originally contained $\rho\in\CC^{k\times k}$ becomes $M_i\rho M_i^{\dagger}$, where $M_i=\sq{P_i}$.
We view selective projections as a generalization of POVMs where the requirement $\sum_i P_i=I$ is replaced by $\sum_i P_i\sdle I$ and the distribution of outcomes is
\m{\PR{\txt{\ord[i] outcome}}
\deq\fr{\tr(P_i\rho)}{\sum_j\tr(P_j\rho)}.}
The class of selective projections is closed \wrt compositions and applying admissible quantum transformations.\fn
{The class of admissible quantum transformations generalizes the class of unitary transformations to include what can be achieved using auxiliary space.}

\sssect{An \hf-state cannot be used twice}

We have seen that each \hf-state can be used to answer at least one \hf-query.
To prove that our \qm\ is secure we will have to argue that an \hf-state cannot be used to answer two complementary \hf-queries with confidence.

Let $\GH$ be the quantum retrieval game corresponding to answering both the possible \hf-queries using one \hf-state, namely
\m{\GH\deq\l(\l(\dr1{16}\dt\kba[x]\r)_{x\in\01^4},\sH\r),}
where
\m{\sH\deq\set[(x,0,a_0,b_0),(x,1,a_1,b_1)\in\hf]{(x,(a_0,b_0,a_1,b_1))}.}
Note that this definition corresponds to the uniform choice of $x\in\01^4$.

\lem[l_hmp_once]{The selective value of $\GH$ is at most $\fr[/]34$.\fn
{From the proof it can be seen that the bound is, actually, tight.}}

\prf{Note that $\sum_x\dr1{16}\dt\kba[x]=\dr14\dt I$.
Consider a selective projection that produces correct answer to $\GH$ with probability $\delta$.
There must exist an answer $(a_0',b_0',a_1',b_1')$ that is produced with non-zero probability, and if it is produced then it is correct with probability at least $\delta$ when $x\sim\U[\01^4]$.
Fix one such answer.

Denote by $E$ the event that $(x,(a_0',b_0',a_1',b_1'))\in\sH$.
By the definition of selective value there exists a projection $P$ such that $\tr(P\rho)>0$ and if the outcome $P$ occurs then $E$ holds with probability at least $\delta$.
We will argue that $E$ cannot be ``witnessed'' very well by any outcome of measuring $\rho$.

Observe that $E$ always corresponds to choosing three different coordinates \f{j_1, j_2, j_3\in[4]} and fixing the values of $x_{j_1}\xor x_{j_2}$ and $x_{j_2}\xor x_{j_3}$.
By symmetry of $\ka[x]$, we can, w.l.g., consider the case of witnessing $x_1=x_2=x_3$ via measuring of $\rho$.

Let $P=\sum_{i\in[k]}\kbra{e_i}$ for some orthonormal $\ket{e_1}\dc\ket{e_k}$.
We have:
\mal{\delta
&\le\fr{\tr\l(P\sum_{x_1=x_2=x_3}\kba[x]\r)}
{\tr\l(P\sum_{x\in\01^4}\kba[x]\r)}
\le\Maxx{i\in[k]}{\fr
{\sum_{x_1=x_2=x_3}\sz{\bket{e_i}{\alpha(x)}}^2}
{\tr\l(\kbra{e_i}\sum_{x\in\01^4}\kba[x]\r)}}\\
&=\fr14\tm\sum_{x_1=x_2=x_3}\sz{\bket{e_0}{\alpha(x)}}^2,}
where $\ket{e_0}\in\set{\ket{e_1}\dc\ket{e_k}}$ attains the optimum of the second inequality (under the ``$0/0=0$'' convention).
Let $e_0^{(j)}\deq\bket{e_0}j$ for $j\in[4]$, then
\mal{\delta
&\le\fr18\l(\l(e_0^{(1)}+e_0^{(2)}+e_0^{(3)}+e_0^{(4)}\r)^2
+\l(e_0^{(1)}+e_0^{(2)}+e_0^{(3)}-e_0^{(4)}\r)^2\r)\\
&=\fr14\l(\l(e_0^{(1)}+e_0^{(2)}+e_0^{(3)}\r)^2
+\l(e_0^{(4)}\r)^2\r)
\le\fr34,}
because $\ket{e_0}$ is a unit vector.}

For $k\in\NN$, let $\GH^k$ be the naturally defined ``product game'' that consists of $k$ independent instances of $\GH$.

\crl[c_GH_prod]{The selective value of $\GH^k$ is at most $\l(\fr[/]34\r)^k$.}

\prf{Let $\GH^{(i)}$ denote the \ord[i] instances of $\GH$ ``inside'' $\GH^k$.
Then
\m[m_chain]{\PR{\GH^k \txt{ is won}}=
\prod_{i\in[k]}\PRr{\GH^{(i)} \txt{ is won}}
{\GH^{(1)}\dc\GH^{(i-1)} \txt{ are won}},}
where the probabilities are defined \wrt the selective measurement being used to play $\GH^k$.

Note that each conditional probability that appears on the right-hand side of \bref{m_chain} is at most $\fr[/]34$:
Otherwise there would exist a selective measurement that used $i-1$ auxiliary instances $\GH^{(1)}\dc\GH^{(i-1)}$, and conditioned upon winning these $i-1$ instances won $\GH^{(i)}$ with probability higher than $\fr[/]34$, contradicting \lemref{l_hmp_once}.
The result follows.}

\ssect[ss_cost]{The cost of counterfeiting a \qm-coin}

Unless stated otherwise, let $\cn$ be the \qm-coin that an adversary is trying to counterfeit, and let $\x$ be the bank's secret string that describes the structure of $\cn$.
We want to argue that in order to achieve his goal, the adversary has to collect certain minimal amount of additional information about the coin, and that task itself is difficult to fulfill.

Let us assume for the rest of our security analysis that the attack under consideration, denoted by $\Ch$, runs at most $U$ instances of \Ver, all of them initiated by sending the identification number of $\cn$.
Informally, $\Ch$ is successful if in the end it outputs quantum states $\rho_1$ and $\rho_2$ (possibly entangled), such that both of them, if given to a trustworthy user, pass \Ver\ with some non-negligible probability.
This probability is viewed \wrt the randomness present in the construction of $\cn$, in $\Ch$ itself, and in the final run of \Ver.

It is crucial that we consider the probability of both the fakes having been accepted.
If instead we were asking what is the smaller of the probabilities that $\rho_j$ passes \Ver\ for $j\in\01$, we would end up with a bound of at least $1/2$:
For example, an adversary can toss $j\sim\U[\01]$ and make $\rho_j$ to be $\cn$, and $\rho_{1-j}$ to be anything.

\lem[l_win_cost]{Consider an attack that completes at most $U$ instances of \Ver\ in order to counterfeit $\cn$.
Conditional upon having passed at most $u\le U$ instances, the success probability of counterfeiting is at most
\m{e^{-\asOm t}+e^{u\ln U-\asOm k}.}
}

\prf{For $j\in\01$, let $I_j$ be a random variable taking the value of the list of \hf-registers that are marked as unused on $\rho_j$.
By the definition of \qm-scheme it should hold that $\sz{I_j}\ge3k/4$ (otherwise the forgery would be obvious right away).

Consider the run of \Ver\ for the counterfeit contained in $\rho_j$.
For any choice of $\Lh$ in step \ref{step_2t/3} and of ``questions'' \set[i\in \Lh]{m_i} in step \ref{step_m}, the quantum measurement applied by the coin holder in step \ref{step_meas} can be decomposed into $\dr{2t}3$ measurements that access individual registers of $\rho_j$ in order to find answers \wrt the corresponding $m_i$.
Let us denote by $P_j^{i,m}$ the measurement applied to $\rho_j$ in order to produce $(a_i,b_i)$ when $m_i=m$.

In step \ref{step_meas} of \Ver\ the holder of $\rho_j$ performs the measurements \set[i\in \Lh]{P_j^{i,m_i}} in order to determine the $\dr{2t}3$ pairs $(a_i,b_i)$ that he will report to the bank.
Now we make two observations that will be crucial for the proof:\itemi{
\item The only pairs of the measurements that do not commute are
\m{\set[{j\in\01, i\in[k], m\in\01}]{\llp P_j^{i,m}, P_j^{i,1-m}\rrp}.}
\item Since the coin holder is now fair to the protocol, the $\dr{2t}3$-set $\Lh$ chosen in step \ref{step_2t/3} of \Ver\ is a uniformly random subset of $I_j$.
The questions $(m_i)_{i\in \Lh}$ are i.i.d.\ by $\U[\01]$.
}

Denote by $V_j$ the instance of \Ver\ that tests $\rho_j$, and accordingly define $\Lh^j$ and $m_i^j$.
Let us view choosing $(m_i^j)_{i\in \Lh^j}$ as first taking $\bm m^j\sim\U[\01^k]$, followed by choosing a random $\Lh^j$ and outputting the projection of $\bm m^j$ to the coordinates in $\Lh^j$.
Clearly, the resulting distribution of $\Lh^j$ and $(m_i^j)_{i\in \Lh^j}$ are the same.
Therefore, we can replace the protocols $V_1$ and $V_2$ by a new quantum procedure, somewhat more friendly to analyze.

Let $\wt V$ be the following procedure that either accepts or rejects quantum states $\rho_1$ and $\rho_2$.\enum{
\item For $j\in\01$, choose $\bm m^j\sim\U[\01^k]$.
\item For $j\in\01$ and $i\in I_j$, apply $P_j^{i,\bm m_i^j}$ to $\rho_j$ and denote the outcome by $(a_i^j,b_i^j)$.
\item For $j\in\01$, choose $\Lh^j$ as a uniformly random subset of $I_j$ of size $\dr{2t}3$.
\item Accept if for all $j\in\01$ and $i\in\Lh^j$ it holds that $(\x_i,\bm m_i^j,a_i^j,b_i^j)\in\hf$; reject otherwise.
}

Observe that all \pl[P_j^{i,\bm m_i^j}] that can appear in a single run of $\wt V$ \e{commute}, and therefore the probability that $\wt V$ accepts exactly equals the probability that both $V_1$ accepts $\rho_1$ and $V_2$ accepts $\rho_2$.

Denote $I\deq I_1\cap I_2$, $I'\deq\set[\bm m^1_i\neq\bm m^2_i]{i\in I}$, $\wt I_j\deq\set[(\x_i,\bm m_i^j,a_i^j,b_i^j)\nin\hf]{i\in I'}$ and $\wt I\deq\wt I_1\cup\wt I_2$.
We will see that $\wt I$ is unlikely to be small, and if it is big then $\wt V$ is unlikely to accept.

Let us first consider the case when the adversary has not run any preliminary protocol and created $\rho_1$ and $\rho_2$ from $\cn$ alone, without any auxiliary knowledge about \x.

By definition, $\sz{I}\ge k/2$.
By uniformity of $\bm m^1$ and $\bm m^2$ it holds that $\E{\sz{I'}}=\sz{I}/2$, and Chernoff bound (\theoref{t_Che}) implies
\m[m_I']{\PR{\sz{I'}\le\fr k5}<e^{-\fr{|I|}{100}}\le e^{-\fr k{200}}.}
By \lemref{l_hmp_once}, for every $i_0\in I'$ it holds that $\PR{i_0\nin\wt I}\le3/4$; moreover, the same remains true even if we condition upon the content of $\wt I\smin\set{i_0}$ (otherwise \lemref{l_hmp_once} would be contradicted by a selective measurement that uses auxiliary instances of $\GH$ in order to win with higher probability, similarly to the proof of \crlref{c_GH_prod}).
Therefore \crlref{c_Che+} can be used here, resulting in
\m[m_It]{\PR{\sz{\wt I}\le\fr{|I'|}5}\le e^{-\fr{|I'|}{200}}.}
Clearly,
\m{\PR{\sz{\wt I}<\fr k{25}}
\le\PR{\sz{I'}\le\fr k5}+
\PRr{\sz{\wt I}\le\fr{|I'|}5}{\sz{I'}>\fr k5},}
which leads, together with \bref{m_I'} and \bref{m_It}, to
\m[m_I_big_no-aux]{\PRr{\sz{\wt I}<\fr k{25}}{\star}
\le e^{-\fr k{200}}+e^{-\fr{k}{1000}},} 
where ``$\star$'' is the condition that $\rho_1$ and $\rho_2$ are created from $\cn$ not using any auxiliary input.

Now assume that in order to produce $\rho_1$ and $\rho_2$ the adversary has completed at most $U$ instances of \Ver, and condition upon at most $u$ of them having been passed successfully.
The idea here is to emulate the same attack, letting the adversary guess the bank's responses locally.
In this form the attack uses no auxiliary data from the bank, which makes $\star$ from \bref{m_I_big_no-aux} hold.

According to \Ver, the only bank's message that depends on $\x$ is the final ``accept''/``reject'' notice.
Therefore, if the adversary (who doesn't know $\x$) does his best to predict all bank's responses, such predictions will be statistically indistinguishable from bank's responses, as long as all ``accept''/``reject'' verdicts are guessed correctly.
The number of different ways to choose at most $u$ ``accepts'' out of $U$ verdicts is at most $U^u+1$, and therefore they are guessed correctly with probability at least $\fr1{U^u+1}$.
Thus from \bref{m_I_big_no-aux},
\m[I_big]{\fr{\PR{\sz{\wt I}<k/5}}{U^u+1}\le e^{-\fr k{200}}+e^{-\fr{k}{1000}}.}

Now assume that $\sz{\wt I}\ge k/5$.
W.l.g., let $\sz{\wt I_1}\ge k/10$.
Then the probability that $\wt V$ accepts is upper-bounded by the probability that none of the elements of $\Lh^1$ comes from $\wt I_1$, and that is at most $(9/10)^{2t/3}<(14/15)^t$.
Together with \bref{I_big}, this implies that
\m{\PR{\txt{$\wt V$ accepts}}<\l(\fr{14}{15}\r)^t
+\l(e^{-\fr k{200}}+e^{-\fr{k}{1000}}\r)\tm(U^u+1),}
as required.}

\ssect[ss_gen_to_near-phased]{Phased attacks}

If we could assume that the attack under consideration is \e{phased}, in a sense that during cheating phase $i$ the \ord[i] steps of all $U$ auxiliary instances of \Ver\ are executed, that would simplify our analysis considerably.
In this part we will show that any attack can be transformed, with a modest loss in the success probability, to the \e{nearly-phased} form.

\ndefi{(phased and nearly-phased attacks)}{Let an attack be using $U$ auxiliary instances of \Ver.

We say that the scenario is phased if it can be viewed as consisting of \ref{step_verdict} consecutive phases, such that at phase $i$ the \ord[i] steps of all $U$ auxiliary instances of \Ver\ are executed.

We call the scenario nearly-phased if it is phased with a relaxation that instead of phases \ref{step_2t/3} and \ref{step_m} it has a phase called ``\ref{step_2t/3} - \ref{step_m}'', when both the \ord[\ref{step_2t/3}] and the \ord[\ref{step_m}] steps of the auxiliary instances of \Ver\ are executed.}

Intuitively, the difference between the two restrictions is that in a nearly-phased scenario an adversary is allowed, say, to choose the $\fr[/]{2t}3$ ``playing'' registers (out of the $t$ suggested by the bank) in the auxiliary instance $1$ of \Ver\ after he has received the $\fr[/]{2t}3$ questions $m_i$ relevant to the auxiliary instance $2$ of \Ver.
In the case of phased attacks such behavior is not allowed:\ the questions $m_i$ relevant to all the auxiliary instances of \Ver\ are available to the adversary only after the choices of ``playing'' registers have been made \wrt all the instances.

The convenience of these definitions comes from the fact that, on the one hand, if an attack is phased then it cannot use in an earlier stage of one auxiliary instances of \Ver\ the output from a later stage of another instances, while on the other hand, only the last bank's response in \Ver\ provides any information about the string \x.
That is, assuming that an attack is (nearly-) phased limits considerably the possibilities for the adversary to use dependencies between different instances of \Ver.

Our claim is the following.

\lem[l_near-phased]{If an attack exists that initiates at most $U$ and wins at least $u\le U$ auxiliary instances of \Ver\ with probability at least $\delta$, then there is a nearly-phased scenario that initiates and completes exactly $U$ and wins exactly $u$ instances of \Ver\ with probability larger than
\m{\fr{\delta-\fr[/]{U}{2^{\fr[/]{2t}3}}}{U^u}.}
}

In the above statement by ``initiating'' an instance of \Ver\ we mean sending a coin identification number to the bank and getting back a list of $t$ registers (step \ref{step_t} of \Ver).

\prf{The proof idea here is somewhat similar to that of \lemref{l_win_cost} -- namely, if the output from a later stage of one auxiliary instance of \Ver\ is used by the adversary in order to decide how to act in an earlier stage of another instance, we would let a ``new adversary'' guess the future response of the first instance before actually receiving it from the bank, and act in the second instance under the assumption that the guessing has been accurate.

Let $\Ch$ be the attack, as guaranteed by the lemma condition.

First of all, let us turn it into $\Ch'$ that always completes $U$ instances of \Ver\ and is likely to win exactly $u$ of them.
This first modification is straightforward -- $\Ch'$ would behave as $\Ch$, except for the following modifications:\itemi{
\item If, according to $\Ch$, no more auxiliary instances are needed but less than $U$ have been run then $\Ch'$ runs ``dummy'' instances of \Ver\ (generating uniformly at random all messages that are sent to the bank), in order to make their total number equal $U$.
\item If, according to $\Ch$, some instances of \Ver\ are aborted, $\Ch'$ completes them in a ``dummy'' way.
\item If at some point it occurs that $\Ch'$ has already won $u$ instances of \Ver, then it completes all remaining instances in a ``dummy'' way.
}
Note that the way $\Ch'$ produces $\rho_1$ and $\rho_2$ is irrelevant for us now, as here we are only interested in the number of ``accepts'' among the preliminary runs of \Ver.

Clearly, the probability that $\Ch'$ wins at least $u$ instances is the same as in the case of $\Ch$; on the other hand, $\Ch'$ wins more than $u$ instances only if at least one ``dummy'' instance has been won.
A single ``dummy'' instance is won with probability exactly $2^{-\fr[/]{2t}3}$, and at least one is won with probability less than $U\tm2^{-\fr[/]{2t}3}$.
Therefore, $\Ch'$ wins exactly $u$ instances of \Ver\ with probability larger than
\m[pr_Ch']{\delta-U\tm2^{-\fr[/]{2t}3}.}

Now let us turn $\Ch'$ into nearly-phased. 
The new attack $\Ch''$ consists of $5$ phases, as follows.
\itemi{
\item[\ref{step_ini}.]
Initiate $U$ instances of \Ver, sending the identification number of $\cn$ to the bank.
Index the instances by $1\dc U$.
\item[\ref{step_t}.]
Get back $U$ \f t-tuples, denoting by $T^{(1)}_i$ the response from the \ord[i] instance of \Ver, $i\in[U]$.
\item[\ref{step_2t/3} - \ref{step_m}.]
Let $W\in\01^U$ be a uniformly chosen binary vector of Hamming weight $u$ -- this is going to be the adversary's guess regarding the winning instances of \Ver.
Start emulating $\Ch'$ skipping the first two steps of each instance of \Ver, as those have been processed already (use \pl[T^{(1)}_i] as bank's responses in step \ref{step_t} of \Ver).
Skip all interaction with the bank beyond step \ref{step_m}; instead, whenever $\Ch'$ acts depending on the bank's final response in the \ord[i] instance of \Ver, emulate $\Ch'$ assuming that the response was $W_i$ (where ``$0$'' stands for ``reject'', and ``$1$'' stands for ``accept'').
Run the emulation until the bank's responses in step \ref{step_m} have been received in all $U$ instances of \Ver.
For $i\in[U]$, denote by $T^{(2)}_i$ the $\fr[/]{2t}3$-tuple chosen in step \ref{step_2t/3} of the \ord[i] instance of \Ver, and by $M_i\in\01^{\fr[/]{2t}3}$ the values sent by the bank in step \ref{step_m} of the \ord[i] instance.
\item[\ref{step_meas}.] Start a new emulation of $\Ch'$, this time skipping steps \ref{step_ini} - \ref{step_m} of each instance of \Ver\ and respectively using the values $T^{(1)}_i$ and $M_i$ as bank's responses.
Do not interact with the bank beyond step \ref{step_meas}; instead, whenever $\Ch'$ acts depending on the bank's final response in the \ord[i] instance of \Ver, emulate $\Ch'$ assuming that the response was $W_i$.
\item[\ref{step_verdict}.] Receive the final responses from the bank, denote them by $V\in\01^U$.
}

It is clear from the construction that $\Ch''$ is nearly-phased.

Let us analyze the probability that $\Ch''$ wins exactly $u$ instances of \Ver.
It is lower bounded by the probability that $V=W$, and that equals the probability that $\Ch'$ wins $u$ instances of \Ver\ and the right $W$ has been guessed in the beginning of phase ``\ref{step_2t/3} - \ref{step_m}'' of $\Ch''$.
The string $W\in\01^U$ is uniformly random of Hamming weight $u$, thus it is correct with probability at least $U^{-u}$.
Therefore, \bref{pr_Ch'} implies that $\Ch''$ wins exactly $u$ instances of \Ver\ with probability larger than
\f{\l(\delta-U\tm2^{-\fr[/]{2t}3}\r)\tm U^{-u},}
as required.}

\ssect[ss_phased_slow]{Phased cheating is slow}

In this section we will prove that nearly-phased attacks require many auxiliary instances of \Ver\ in order to win enough of them for \lemref{l_win_cost} to allow non-negligible counterfeiting success probability.

\lem[l_phased_slow]{A nearly-phased attack that initiates and completes $U$ auxiliary instances of \Ver\ wins at least $\fr[/]{3k}t$ of them with probability at most
\m{e^{2\ln U-\asOm{\fr[/]{t^2}k}}.}}

As before, by ``initiating'' an instance of \Ver\ we mean sending a coin identification number to the bank and getting back a list of $t$ registers (step \ref{step_t} of \Ver).

\prf{Let $\Ch$ be the nearly-phased attack under consideration.
For $i\in[U]$, let random variables $T^{(1)}_i$, $T^{(2)}_i$ and $M_i$ describe the transcript of the \ord[i] instance of \Ver, as follows:\itemi{
\item $T^{(1)}_i$ takes the value of the $t$-tuple chosen by the bank in step \ref{step_t};
\item $T^{(2)}_i$ takes the value of the $\fr[]{2t}3$-tuple chosen by the adversary in step \ref{step_2t/3};
\item $M_i\in\01^{\fr[/]{2t}3}$ contains the $\fr[]{2t}3$ ``questions'' chosen by the bank in step \ref{step_m}.
}
For $j\in T^{(1)}_i$, let $T^{(1)}_i[j]$ be the position of $j$ in $T^{(1)}_i$, and similarly define $T^{(2)}_i[j]$.
For $j\in T^{(2)}_i$, let $M_i[j]$ be the \ord[{T^{(2)}_i[j]}] bit of the value received by $M_i$ -- that is, $M_i[j]$ denotes the \hf-query asked in the \ord[i] instance of \Ver\ \wrt the register $j$.

For $i,j\in[U]$, $i\ne j$, let $S^{(2)}_{i,j}\deq T^{(2)}_i\cap T^{(2)}_j$ (viewed as a set) and
\m[m_St]{\wt S^{(2)}_{i,j}
\deq\set[s\in S^{(2)}_{i,j}]{M_i[s]\neq M_j[s]}.}
That is, $S^{(2)}_{i,j}$ contains the registers of \cn\ that are part of the bank's challenge questions both in the \ord[i] and in the \ord[j] auxiliary instances of \Ver, and $\wt S^{(2)}_{i,j}$ contains the registers where good answers to the both possible \hf-queries have to be found in order to pass both the \ord[i] and the \ord[j] instances of \Ver.
Note that the attack $\Ch$ produces answers to all the relevant \hf-queries not having any auxiliary information about \x\ ($\Ch$ is phased, and the only bank's responses that contain information about \x\ are the final ones, which were not available to $\Ch$ at the earlier phase).

Denote by $\Cl W_i$ the event that the \ord[i] instance of \Ver\ is won, and by $\Cl W_{i,j}$ the event that both the \ord[i] and the \ord[j] instances are won.
For every $i\ne j$, \crlref{c_GH_prod} implies that
\m[m_Wij]{\PRr{\Cl W_{i,j}}{T^{(2)}_1\dc T^{(2)}_{[U]},M_1\dc M_{[U]}}\le
\l(\fr[/]34\r)^{\sz{\wt S^{(2)}_{i,j}}}.}

Let $\wt r\in\NN$, and denote by $\wt{\Cl E}$ the event that $\Cl W_{i,j}$ does not hold whenever $\sz{\wt S^{(2)}_{i,j}}\ge\wt r$ (later we will fix $\wt r$ to make $\Cl W_{i,j}$ very likely to occur).
Then from \bref{m_Wij},
\m[m_Et]{\PR{\wt{\Cl E}}\ge1-U^2\tm\l(\fr[/]34\r)^{\wt r}.}

Similarly, let $r\in\NN$ (to be fixed later), and denote by $\Cl E$ the event that $\Cl W_{i,j}$ does not hold whenever $\sz{S^{(2)}_{i,j}}\ge r$.
Let $\Cl E'$ be the event that $\sz{\wt S^{(2)}_{i,j}}\ge\wt r$ whenever $\sz{S^{(2)}_{i,j}}\ge r$, then from \bref{m_Et},
\m[m_E]{\PR{\Cl E}\ge\PR{\Cl E'}-U^2\tm\l(\fr[/]34\r)^{\wt r}.}

When $\Cl E$ holds, any two different elements of the family
\m{\Cl F\deq\set[\Cl W_i \txt{ holds}]{T^{(2)}_i}}
share less than $r$ elements.
We choose
\m{r\deq\floor{\fr{2t^2}{9k}},}
then \lemref{l_sets} implies that $\sz{\Cl F}<\fr[/]{3k}t$, i.e.,
\m[m_if_E]{\txt{$\Cl E$ holds}
\tb\implies\tb
\txt{Less than $\fr[/]{3k}t$ instances of \Ver\ are won}.}

It remains to show that $\Cl E$ is likely to hold.
Fix
\m{\wt r\deq\floor{\fr{t^2}{10k}},}
and let us see that $\Cl E'$ is very likely to occur.

Before we deal with the nearly-phased case, suppose that $\Ch$ is phased.
In this case there is no dependence between the variables $\l(T^{(2)}_i\r)_{i=1}^U$ and the variables $\l(M_i\r)_{i=1}^U$, and therefore $\wt S^{(2)}_{i,j}$ is a randomly chosen subset of $S^{(2)}_{i,j}$, where each $s\in S^{(2)}_{i,j}$ independently becomes an element of $\wt S^{(2)}_{i,j}$ with probability $\fr[/]12$.
By Chernoff bound (\theoref{t_Che}), if $\sz{S^{(2)}_{i,j}}\ge r$ then $\PR{\sz{\wt S^{(2)}_{i,j}}<\wt r}\le e^{-\asOm r}$, and therefore
\m[m_E']{\PR{\Cl E'}\ge1-U^2\tm e^{-\asOm r}.}

When $\Ch$ is nearly-phased, the variables $\l(T^{(2)}_i\r)_{i=1}^U$ are not necessarily independent from $\l(M_i\r)_{i=1}^U$ (the adversary is allowed to choose the $\fr[/]{2t}3$ ``playing'' registers in step \ref{step_2t/3} of the \ord[i] instance of \Ver, depending on $M_j$ received from the bank in step \ref{step_m} of the \ord[j] instance of \Ver, $j\neq i$).
However, we claim that for every $j\neq i$ the values $\set[s\in T^{(2)}_i\cap T^{(2)}_j]{M_i[s]\xor M_j[s]}$ are unbiased and mutually independent -- and this is all we need for \bref{m_E'} to hold (cf.\ \bref{m_St}).
Indeed, from the definition of \Ver\ it is clear that at least one of $M_i$ and $M_j$ is chosen by the bank uniformly at random after the values of both $T^{(2)}_i$ and $T^{(2)}_j$ have been set by the choice of the adversary, and therefore $M_i[s]\xor M_j[s]$ is unbiased.

From \bref{m_E} and \bref{m_E'},
\m{\PR{\Cl E}\ge1-e^{2\ln U-\asOm{\fr[/]{t^2}k}}.}
Together with \bref{m_if_E}, this implies the result.}

\ssect[ss_together]{\qm\ is safe}

We are ready to prove the main theorem.

\theorep{t_main}{\teoMain}

\prf{From \lemref[l_near-phased]{l_phased_slow} it follows that an attack $\Ch$ that receives the first responses in at most $U$ auxiliary instances of \Ver\ can win at least $\fr[/]{3k}t$ of them with probability at most
\m{e^{\asO{\fr[/]kt}\ln U-\asOm{\fr[/]{t^2}k}}.}
Then \lemref{l_win_cost} implies that $\Ch$ succeeds in counterfeiting the coin $\cn$ with probability at most
\m{e^{\asO{\fr[/]kt}\ln U-\asOm{\fr[/]{t^2}k}},}
and the result follows.}

\sect[s_opti]{Optimality of \qm}

In this part we consider a generic quantum money scheme with classical verification, where the qubit-size of a coin is $K$ and a secret bank record describing a coin contains $R$ bits.

Let us define the \e{counterfeiting complexity} of a quantum money scheme as
\m{\Minn{\eps}{\Max{\fr[/]1{\eps}, \lla\txt{time required to counterfeit a coin with success probability at least $\eps$}\rra}},}
this definition is a lower bound on what we intuitively mean by ``time required to forge a counterfeit''.\fn
{Instead, one might consider the time required to counterfeit a coin with \e{constant} success probability.
The (asymptotic) time complexity of an attack that succeeds with constant probability is an upper bound on the counterfeiting complexity, as defined above.
Note that our scheme from the previous section has high counterfeiting complexity, therefore it is secure in the stronger sense.
On the other hand, the upcoming (formal) optimality statements will be made \wrt attacks that achieve constant success probability, which will make those statements also as strong as possible.
Intuition-wise, we find the definition with ``flexible $\eps$'' more appealing, that is why we use it in the informal discussion.}
Note that \theoref{t_main} and \crlref{c_main} imply that the counterfeiting complexity of \qm\ is exponential both in $K$ and in $R$.

First of all, $2^R$ adversarial verification attempts are enough to exhaustively check all possible bank's records, and therefore \asO{2^R} is an upper bound on the counterfeiting complexity of any quantum scheme.
So, in the case of \qm\ the length of a bank record as a function of counterfeiting complexity is polynomially-close to optimal. 

Can the counterfeiting complexity be super-exponential in $K$?
We could not find a simple argument against this possibility.
The counterfeiting complexity of \qm\ is exponential in $K$ (which can probably be viewed as ``reasonably good''), and we leave the question above as an open problem.

There is one parameter in the construction of \qm\ that one might like to improve -- namely, the number of verification rounds that a new quantum coin can go through before it has to be returned to the bank.
In this section we show that no scheme can allow this number to be larger than linear in $K$, and therefore our construction is polynomially-close to the optimal in this respect also.

\theo[t_opti]{Let $T$ be the number of times that a new coin can be verified via classical communication with the bank before it has to be replaced.
Suppose that if a fair user verifies a fresh coin $T$ times in a sequence then all $T$ verifications are passed with probability at least $\dr89$.
Then either $T$ auxiliary instances of the verification protocol are sufficient for an adversary to counterfeit a coin with probability at least $\dr23$, or a coin contains \asOm{T} qubits (or both).}

To prove the theorem we will need the following technical statement (which might be of independent interest).

\lem[l_mut]{Let $A$ and $B$ be discrete random variables, such that there exists a condition that can be satisfied with probability at most $\alpha$ by the value of any random variable independent from $A$.
If the value of $B$ satisfies the condition with probability at least $\beta\ge\alpha$, then
\m{I(A:B)\ge2(\beta-\alpha)^2.}}

First we prove the theorem, then the lemma.

\prfsk[\theoref{t_opti}]{Assume that more than $T$ auxiliary instances of the verification procedure are required for an adversary to counterfeit a coin with probability at least $\dr23$.

To argue that a coin consists of \asOm{T} qubits, let us show that its ``quantum part'' has mutual information \asOm{T} with bank's secret record.
To make sure that we are not counting information carried by the classical part of a coin, let us assume \wlg that the first message of the verification protocols is sent by the coin holder to the bank and contains all the classical information that the coin contained when it was fresh.

Let $L_1\dc L_T$ be random variables, respectively taking values of the transcripts of $T$ sequentially executed protocols for coin verification via classical communication (assuming that the coin holder fairly follows the protocol, and that the coin was fresh when the first verification started).
For convenience (and w.l.g.), assume that a transcript provides complete information about the action taken by a fair coin holder \wrt the coin being verified.
Let $\rho$ be the mixed state of a fresh coin whose secret record is unknown, and for every $j\in[T]$ and $\ell_1\dc\ell_j$, let $\rho_{\ell_1\dc\ell_j}$ be the state of a fresh coin that went through $j$ verification protocols whose transcripts were, respectively, $\ell_1\dc\ell_j$.

Let $(\ell_i)_{i=1}^T$ be the values taken by $(L_i)_{i=1}^T$.
Denote by $\Cl R$ a random variable describing the bank's secret record corresponding to the coin under consideration.
Let $S(\dt,\dt)$ denote quantum mutual information, we claim that
\m[m_opt_ind]{\forall i\in[T]:
\E{S(\rho_{\ell_1\dc\ell_{i-1}}:\Cl R)
-S(\rho_{\ell_1\dc\ell_i}:\Cl R)}
\ge\fr1{2592},}
where the expectation is taken \wrt the choice of $(L_i)_{i=1}^T$.
From \bref{m_opt_ind} it follows that $S(\rho:\Cl R)\in\asOm{T}$, and Holevo's bound implies that $\rho$ consists of \asOm{T} qubits, as required.

To prove \bref{m_opt_ind} we will also use Holevo's bound.
For simplicity let us assume that each run of the verification protocol requires exactly one quantum measurement to be performed by the coin holder (the case of many measurements is treated similarly).

Fix $i\in[T]$.
Observe that if during the \ord[i] run of the verification protocol the coin holder would not perform any quantum measurement, instead making the ``best guess'' responses  based on the previous transcripts $\ell_1\dc\ell_{i-1}$, then the probability to pass the verification would be less than $\dr78$ -- otherwise an adversary could, based on the transcripts $\ell_1\dc\ell_{i-1}$, prepare two counterfeits that would both pass the verification with probability at least $(\dr78)^2>\dr23$, contradicting the assumptions of the theorem.
On the other hand, by making the quantum measurement, as prescribed by the verification protocol, a fair coin holder is able to pass the verification with probability at least $\dr89$, also guaranteed by the theorem assumptions.

Conditional on $\ell_1\dc\ell_{i-1}$ being the values taken by $L_1\dc L_{i-1}$, the following holds.
The acceptance condition of the \ord[i] verification can be satisfied with probability at most $\dr78$ by a random variable that doesn't depend on $\Cl R$; at the same time the outcome of the quantum measurement performed by the coin holder satisfies the condition with probability at least $\dr89$.
Therefore, \lemref{l_mut} implies that the expected conditional mutual information between the measurement outcome and $\Cl R$ is at least $\dr1{2592}$.
Holevo's bound implies \bref{m_opt_ind}, and the result follows.}

\prf[\lemref{l_mut}]{W.l.g., assume that the condition under consideration is a function of the value taken by $B$.
Let $X$ and $Y$ be the supports of $A$ and $B$, respectively.
For every $b\in Y$, let $X_b\sbseq X$ be the set of values of $A$ that satisfy the condition when $B=b$.
Let $\mu$ be the distribution of $A$, and let $\mu_b$ be the distribution of $A$ conditional upon $B=b$.
Let $\alpha_b\deq\mu(X_b)$ and $\beta_b\deq\mu_b(X_b)$.
The requirements of the lemma assure that
\m[m_lem_as]{\E[B=b]{\alpha_b}\le\alpha
\tbb\txt{and}\tbb
\E[B=b]{\beta_b}\ge\beta.}

By definition,
\m[m_I_def]{I(A:B)=\E[B=b]{\KL{\mu_b}{\mu}}
=\E[B=b]{\sum_x\mu_b(x)\tm\log\l(\fr{\mu_b(x)}{\mu(x)}\r)},}
and this is the value we want to bound from below.
We have
\mal{\sum_{x\in X_b}\mu_b(x)\tm\log\l(\fr{\mu_b(x)}{\mu(x)}\r)
&=\alpha_b\sum_{X_b}\fr{\mu_b(x)}{\alpha_b}
\tm\log\l(
\fr{\dr{\mu_b(x)}{\alpha_b}}{\dr{\mu(x)}{\beta_b}}
\tm\fr{\alpha_b}{\beta_b}\r)\\
&=\alpha_b\tm\l(\KL{\fr{\mu_b}{\alpha_b}}{\fr{\mu}{\beta_b}}
+\log\l(\fr{\alpha_b}{\beta_b}\r)\r)
&\ge\alpha_b\tm\log\l(\fr{\alpha_b}{\beta_b}\r),}
where the inequality follows from non-negativity of $\KL{\dt}{\dt}$ and the fact that, restricted to $X_b$, both $\dr{\mu_b}{\alpha_b}$ and $\dr{\mu}{\beta_b}$ are probability distributions.
Similarly,
\m{\sum_{x\nin X_b}\mu_b(x)\tm\log\l(\fr{\mu_b(x)}{\mu(x)}\r)
\ge(1-\alpha_b)\tm\log\l(\fr{1-\alpha_b}{1-\beta_b}\r),}
leading to
\m{\KL{\mu_b}{\mu}
\ge\alpha_b\tm\log\l(\fr{\alpha_b}{\beta_b}\r)
+(1-\alpha_b)\tm\log\l(\fr{1-\alpha_b}{1-\beta_b}\r)
=\KL{\Cl B_{\alpha_b}}{\Cl B_{\beta_b}}
\ge2(\alpha_b-\beta_b)^2,}
where $\Cl B_p$ denotes Bernoulli distribution with probability $p$ for outcome ``1'', and the last inequality follows from Pinsker's inequality.
Plugging this into \bref{m_I_def}, we obtain the desired
\m{I(A:B)\ge2\E[B=b]{(\alpha_b-\beta_b)^2}
\ge2(\beta-\alpha)^2,}
where the last inequality follows from \bref{m_lem_as}.}

\sect{Conclusions}

We constructed a quantum money scheme \qm\ that allows verifying a coin via classical communication with a bank.
Thus we are proving existence of secure quantum money schemes that do not require quantum communication for coin verification.

Our scheme has the following properties.
\itemi{
\item The coins are exponentially hard to counterfeit, even if an adversary is adaptively using repeated verification attempts in order to collect information about a coin.
\item The classical communication channel used for verification can be unsecured.
\item The database of the bank is static.
\item The dependence between the number of verifications that a \qm-coin can go through and the number of qubits that it contains is optimal, up to a polynomial.
}

There are (at least) two questions that remain open:
\itemi{
\item Is it possible to build \e{anonymous} quantum money schemes with classical verification, by allowing multiple identical instances of quantum coins, as suggested by Mosca and Stebila~\cite{MS10_Qu_Co}? 
\item Is it possible to have the counterfeiting complexity of quantum money super-exponential in the number of qubits that a coin contains (cf. \sref{s_opti})?
}

\subsection*{Acknowledgments}
I am grateful to Scott Aaronson and Martin R{\"o}tteler for numerous helpful discussions.
Dana Moshkovitz has drawn my attention to the result of~\cite{IK10_Con}.
I have received many helpful comments from Ronald de Wolf, and various comments from anonymous reviewers, mostly helpful.
I acknowledge support by ARO/NSA under grant W911NF-09-1-0569.

\bibliography{tex}

\MyComment{Spell-check}

\end{document}